\documentclass[epjST]{svjour}
\usepackage{graphics,amssymb,amsmath,bm,color}

\newcommand{\rf}[1]{(\ref{#1})}

\begin{document}
\title{Transitions in large eddy simulation of box turbulence}
\author{Lennaert van Veen\thanks{\email{Lennaert.vanVeen@uoit.ca}}\fnmsep\inst{1} \and
  Genta Kawahara\inst{2} \and Tatsuya Yasuda\inst{3}}
\institute{University of Ontario Institute of Technology, Oshawa, Ontario, Canada \and 
  Osaka University, Toyonaka, Osaka, Japan \and Imperial College, London, United Kingdom}
\abstract{
One promising decomposition of turbulent dynamics is that into building blocks such as equilibrium and periodic solutions and 
orbits connecting these. While the numerical approximation of such building blocks is feasible for flows in small domains
and at low Reynolds numbers, computations in developed turbulence are currently out of reach because of the large number of degrees of 
freedom necessary to represent Navier-Stokes flow on all relevant spatial scales. We mitigate this problem by applying large eddy 
simulation (LES), which aims to model, rather than resolve, motion on scales below the filter length, which is fixed by a model parameter. 
By considering a periodic spatial domain, 
we avoid complications that arise in LES modelling in the presence of boundary layers. We consider the motion
of an LES fluid subject to a constant body force of the Taylor-Green type as the separation between the forcing length scale and the filter length
is increased. In particular, we discuss
the transition from laminar to weakly turbulent motion, regulated by simple invariant solution, on a grid of $32^3$ points.
} 
\maketitle
\section{Introduction}
\label{intro}

From the point of view of dynamical systems theory, the essential difficulties of turbulence research are the {\em spatio-temporal complexity} and the {\em irreproducibility}
of turbulent flow. Because of the complexity, it is a hard task to extract from snap shots of a turbulent flow field, whether obtained experimentally or through 
numerical simulation, the essential dynamical processes that sustain turbulence. The irreproducibility, due to the extremely sensitive dependence on initial conditions,
makes it hard to establish the relevance of such processes if they can be identified. One way to attack these problems is to consider 
the Simple Invariant Solutions (SIS) that the governing equations admit, even if the unconstrained flow, observed when starting from random initial conditions,
is highly turbulent. The relation between these invariant solutions, such as periodic
orbits and travelling waves, and turbulence is not clear {\em a priori}. They are, however, reproducible and their spatio-temporal dynamics can be investigated in
great detail. 

One way to query the relevance of SIS is to compute some 
physically
interesting time-mean quantities, such as velocity profiles or energy spectra, along the SIS and compare them to those of the surrounding turbulence.
If the mean quantities of some particular invariant solution are close to those of turbulence, we may assume that this solutions is a good ``model'' of the turbulent motion. 
We then assert that the dynamical processes observed in the SIS are relevant to the turbulent dynamics.
An early examples of work along these lines can be found in Kawahara and Kida \cite{KK} and 
van Veen {\sl et al.} \cite{VKK}. A similar, but more sophisticated, idea is to consider a set of invariant solutions as a whole to model turbulence.
The turbulent state is then thought to wander among the invariant solutions, following their stable and unstable manifolds closely. Examples of this strategy include the
description of marginally turbulent Couette flow by Halcrow {\sl et al.} \cite{Halcrow} and the work of Lucas and Kerswell on two dimensional
Kolmogorov flow \cite{LK}. While the first approach is more practical, given the difficulty of computing SIS explained below, the latter has 
a theoretical basis in periodic orbit theory \cite{POT}.

Whichever way we decide to compose a model of turbulence out of SIS, the technical problem to solve is: how to compute them? 
Because of the spatio-temporal complexity of turbulence, any numerical representation of the state of the fluid will require a large number of Degrees Of Freedom (DOF).
The SIS we seek to compute are then solutions to well-posed Boundary Value Problems (BVPs) in space and time for these DOF, supplemented by 
parameters like the speed of propagation for travelling waves and the period for periodic orbits.
These BVPs can only be solved by iterative methods, usually chosen from a family of modified Newton iteration methods. The current state of the art for implementing 
such Newton-like methods uses inexact linear solvers and can be summarised as a ``triple loop'' (see, e.g. S\'anchez and Net \cite{SanchNet}).
The outer loop is over (modified) Newton-Raphson iterations
for solving the BVP. The second loop is over Krylov iterations, refining the Newton update step, which is the solution to system of linear equations that involves the
Jacobian matrix of the BVP. Each Krylov iteration requires computing the product of this matrix with a given perturbation vector. To this end, we time-step the fluid simulation
along with the linearised equations along the approximate solution in the inner loop. Thus, the time it takes to compute a SIS is proportional to the product of 
the number of (modified) Newton iterations, the number of Krylov steps, the number of time steps and the wall time per step. Each of these factors is sensitive to the Reynolds
number of the ambient turbulent flow. We will demonstrate this by a few coarse estimates for the case of Homogeneous Isotropic Turbulence (HIT).

Starting with the inner loop, and assuming we use a pseudo-spectral time-stepping code on a Fourier basis, as has been the standard for HIT since the pioneering work of
Orszag and Patterson \cite{OrsPat}, we can estimate the computational complexity as follows. The cost of a single time step is proportional to $[n\log(n)]^3$ if $n$ is the 
linear grid size, or, equivalently, the number of Fourier modes needed to resolve the dependence of one scalar variable on one spatial coordinate. We adopt the commonly
used estimate for the ratio between the largest and smallest spatial scales in HIT, $L/\eta\propto Re^{3/4}\propto Re_{\lambda}^{3/2}$, where $\eta$, the Kolmogorov scale,
approximates the smallest scale of coherent eddies and $L$, the integral scale, denotes the largest and $Re_{\lambda}$ is Taylor's microscale Reynolds number (see, e.g., Constantin {\sl et al.}
\cite{CFMT}).
Taking a sufficiently fine grid, and dropping the logarithmic correction, we thus find that a single time step contributes a factor of $Re_{\lambda}^{9/2}$ to the overall 
complexity. The number of time steps must be chosen such that processes on the scale $\eta$ are resolved properly. Since the Kolmogorov time scale scales with 
$Re_{\lambda}$ as $\tau_{\eta}\propto Re_{\lambda}^{-1}$, the number of time steps will grow in proportion to $Re_{\lambda}$. The time-stepping thus contributes a factor
of $Re_{\lambda}^{11/2}$ to the overall complexity. As for the number of Krylov steps, from the work of S\'anchez et al. \cite{Sanch} we know that Krylov-type methods
efficiently resolve high-wave number modes that are severely damped by viscosity. Perturbations that grow or are only mildly damped in time need to be resolved by iterations
of the algorithm. It then seems reasonable to assume that the number of Krylov iterations grows in  proportion to $n^3\propto Re_{\lambda}^{9/2}$ in the high Reynolds number limit. Putting these factors together,
we find that the computational cost of performing a single (modified) Newton step will grow as $Re_{\lambda}^{10}$. Depending on the forcing mechanism, the transition to
turbulence 
is usually observed for $Re_{\lambda}$ of order $10$, and here computations are relatively painless on grids of up to $64^3$ points. In order to separate the
energy containing and Kolmogorov length scales, and see a power-law scaling of the energy spectrum, $Re_{\lambda}$ must be about ten times larger. The prospect of the computations
taking a factor of $10^{10}$ more time seems to preclude the application of the SIS-based decomposition of HIT.

Obviously, one can mitigate the problem by performing the computations in parallel where possible. Since the Krylov solving is an inherently sequential process, 
the parallelism is usually in the time-stepping, in particular in the Fast Fourier Transform (FFT). Since the parallel FFT requires a lot of communication, its
efficiency depends sensitively on the computer architecture and concurrent usage, but in our experience the scaling can be nearly linear for a number of processes up
to $n/2$. Further speed-up can be achieved by moving to GPU processing. Depending on the details of the GPU cards and the implementation, this could be several times faster than
conventional CPU parallelism. Assuming we have unlimited access to a great many CPUs/GPUs, we may thus speed up the computations by a factor proportional to $Re_{\lambda}^{3/2}$
or so, making only a small sent in the frightful estimate of $Re_{\lambda}^{10}$.

In the current work, we explore the possibility of using Large Eddy Simulation (LES) as a partial solution to the problem sketched above. In LES, a spatial filter is applied
to the equations of fluid motion, effectively quenching motion on scales smaller than a cut-off fixed by a parameter we can choose freely. Since the small-scale dynamics is no
longer resolved, its influence on the fluid motion must be modelled. For this end, we employ a closure due to Smagorinsky \cite{Smagorinsky}. With the resulting model, we can 
approximate the motion of the fluid on scales on which the inertial terms dominate without spending computational resources on resolving motion strongly damped by viscosity. 
The goal of LES is to model flow at infinite Reynolds number with a finite number of DOF. We can fix the size of the computational grid arbitrarily, thereby fixing the 
ratio of the largest to the smallest scales of coherent motion. Consequently, we can study turbulent flow on grids as small as $32^3$ or $64^3$ and greatly reduce the cost of
time stepping in the inner most loop. 

The freedom to fix the grid size comes at a price. Most importantly, we are no longer studying Navier-Stokes flow, but rather the dynamics of a hypothetical ``LES fluid'' with
a highly nontrivial stress-strain relation. The relevance of LES dynamics to real turbulence is not undisputed and depends on the quantities of interest. Lesieur and Matais
note that Smagorinsky's closure is ``The most popular model for engineering application purposes when small-scale turbulence is three dimensional (\ldots)'' and that it 
``(\ldots) gives interesting results (as far as coherent structures are concerned, for instance) in isotropic turbulence and in free-shear flows (\ldots)'' \cite{Lesieur}. 
Indeed, coherent structures
and their dynamics in turbulence without material boundaries are our primary interest in this study, and thus it seems reasonable to use the basic Smagorinsky model rather than more sophisticated closure models 
that introduce additional nonlinearities in the equations of motion.

We study the Smagorinsky LES fluid on a three-dimensional, cubic domain with periodic boundary conditions and steady external forcing. The forcing takes the form of four counter rotating
vortex columns and the resulting laminar flow resembles one a a family of flows introduced by Taylor and Green \cite{TG}. We fix the grid size to $32^3$ points, resulting in a dynamical
system with $28,484$ DOF which can easily be time-stepped on a modern desktop computer with four CPUs. The only control parameter is the Smagorinsky constant, which effectively
sets the filter length. If the filter length is several times greater than the grid spacing, the LES is ``over damped'' or ``redundant'', meaning we resolve motion on spatial scales
on which the eddy viscosity is dominant. In this regime, the laminar flow is a stable equilibrium. We track this equilibrium solution to smaller values of the Smagorinky constant
and compute its first and second bifurcation. From the first, a branch of periodic orbits emerges while from the second a family of equilibria is born. We compute several instabilities of
these secondary branches and the tertiary branches of periodic orbits that they produce. The SIS we compute this way are called {\em derivative solutions} as
they are all connected to the laminar flow through a sequence of bifurcations. We show, that the derivative solutions look very different from the ambient turbulence. Firstly, they all
inherit several discrete symmetries from the laminar flow and secondly, they have amplitudes much smaller than typical turbulent fluctuations.

A different strategy to find SIS 
is to ``filter'' approximately invariant solutions from a turbulent time series. We parse the time series and detect segments that resemble equilibrium and time-periodic solutions.
If found, we feed the corresponding initial conditions to a globally convergent Newton iteration scheme. In particular, we use the Newton-hook step, integrated with the Newton-Krylov
algorithm as proposed by Viswanath \cite{Vis}.
The term ``globally convergent'' is a misnomer; while this
Newton iteration scheme has a much larger radius of convergence than conventional Newton-Raphson iteration, there is no guarantee of convergence. 
The rate of success tends to be high for weakly turbulent flows and rapidly decay as the Reynolds number increases. We apply the filtering method
in an over damped regime where neither the laminar flow nor the derivative solutions are linearly stable, and identify an orbit of long period and large amplitude.

The long-period, large-amplitude SIS is computed at a filter length about ten times smaller than the integral scale $L$. At this scale separation, the smallest scale
motion is still strongly influenced by the external forcing and is thus neither homogeneous nor isotropic. In ongoing work, we are aiming to compute silimar SIS
representative of LES turbulence with a larger scale separation. Given the estimates above for the computational complexity of the Newton-Krylov-hook apporach, this 
will surely be a time-consuming process of trial and error, to which the current results present a promising point of entry.

\section{Taylor-Green flow with the Smagorinsky model}
\label{sec:model}
The incompressible Navier-Stokes equation with periodic boundary conditions and body forcing is given by
\begin{align}\label{NS}
\bm{u}_t+\bm{u}\!\cdot\! \nabla \bm{u}+\frac{1}{\rho}\nabla p-\nu \Delta \bm{u}&=\gamma \bm{f} \\
\nabla\!\cdot\! \bm{u}&=0\nonumber\\
\bm{u}(x+L,y,z,t)=\bm{u}(x,y+L,z,t)=& \nonumber\\
\bm{u}(x,y,z+L,t)=\bm{u}(x,y,z,t)&\nonumber
\end{align}
where $\rho$ is the fluid density, assumed to be constant, $\nu$ is the kinematic viscosity and $\gamma$ is the amplitude of the forcing.

We will numerically approximate solutions to system \rf{NS} on a regular grid of $n^3$ points. In order to reduce the number of degrees of 
freedom in the numerical simulations, we adopt the strategy proposed by Smagorinsky \cite{Smagorinsky} 
to model the model the effect of motion on the smallest scales with an effective eddy viscosity. This approach can be interpreted as
the application of a spatial filter given by a convolution
\begin{equation}\label{filter}
\bar{\bm{u}}=G\ast \bm{u}
\end{equation}
The kernel $G$ can be thought of as a low-pass filter with width $l_{\rm f}$. Since the spatial filter commutes with all spatial derivatives, the treatment of the linear
terms in system \rf{NS} is straightforward.
Filtering the advection term we obtain
\begin{equation}\label{Smodel1}
\overline{u_i\partial_i u_j}=\partial_i\overline{u_iu_j}=\partial_i(\bar{u}_i\bar{u}_j)+\partial_i(\overline{u_iu_j}-\bar{u}_i\bar{u}_j)\equiv \bar{u}_i\partial_i \bar{u}_j+\partial_i \tau^{({\rm s})}_{ij}
\end{equation}
where summation over repeated indices is implied and the last equation defines the {\em sub grid stress} tensor $\tau^{({\rm s})}$. It is split into a diagonal component, that modifies the pressure, and a deviatoric component, according to
\begin{equation}\label{Smodel2}
\tau^{({\rm s})}_{ij}=\frac{1}{3}\Pi\delta_{ij}+2\nu_T \bar{S}_{ij}
\end{equation}
where $\bar{S}=(\partial_i \bar{u}_j+\partial_j \bar{u}_i)/2$ is the filtered rate-of-strain tensor, $\delta$ is the Kronecker delta symbol and the {\em eddy viscosity} $\nu_T$ is defined by
\begin{equation}\label{Smodel3}
\nu_T=(C_{\rm S} \Delta)^2 \sqrt{2\bar{S}_{ij}\bar{S}_{ij}}
\end{equation}
This expression for $\tau^{({\rm s})}$, which contains a filtered term quadratic in $u$, in terms of the filtered rate of strain closes the system of equations and constitutes the Smagorinsky model.
The closure introduces a dimensionless parameter, $C_{\rm S}$, called the {\em Smagorinky parameter}.

The resulting equations are
\begin{align}\label{Smodel4}
\partial_t\bar{\bm{u}}+\bar{\bm{u}}\!\cdot\!\nabla \bar{\bm{u}} +\nabla \left(\frac{\bar{p}}{\rho}+\frac{1}{3}\Pi\right) -2\nabla([\nu_T+\nu]\bar{\bm{S}})&=\gamma \bar{\bm{f}} \\
\nabla\!\cdot\! \bar{\bm{u}}&=0
\end{align}
We fix $\bar{\bm{f}}=(-\sin(k_{\rm f}x)\cos(k_{\rm f}y),\cos(k_{\rm f}x)\sin(k_{\rm f}y),0)^t$ with $k_{\rm f}=2\pi/L$. This body force induces a flow with four counter-rotating vortex columns,
which is one of a family of flows studied by Taylor and Green\cite{TG}.

In the following, we nondimensionalize the equations according to
\begin{alignat}{3}\label{nondim}
\bm{x}'&=k_{\rm f}\,\bm{x}      &\qquad t'&=\sqrt{L\gamma k^2_{\rm f}}\,t &\qquad \Pi'&=\frac{1}{L\gamma}\,\Pi \\
\bar{\bm{u}}'&=\sqrt{\frac{1}{L\gamma}}\,\bar{\bm{u}} &\qquad \bar{p}'&=\frac{1}{\rho L\gamma}\,\bar{p} & &
\end{alignat}
and drop both the over bars for spatially filtered quantities and the primes for non dimensional quantities. The resulting momentum balance equation is
\begin{equation}\label{nondimLES}
\bm{u}_t+\bm{u}\!\cdot\! \nabla \bm{u}+\nabla \left(p+\frac{1}{3}\Pi\right)-2\nabla \left(\left[\frac{1}{R^{3/2}}\nu_T +\frac{1}{Re} \right]\bm{S}\right)=\frac{1}{\alpha} \bm{f} 
\end{equation}
with non dimensional parameters
\begin{alignat}{3}\label{Reyal}
R&=\left(\frac{1}{C_{\rm S}\Delta k_{\rm f}}\right)^{4/3} &\qquad Re&=\frac{\sqrt{L\gamma}}{\nu k_{\rm f}} &\qquad \alpha&=L k_{\rm f}=2\pi
\end{alignat}
Here, $R$ is the ``LES Reynolds number'', which measures the ratio of the forcing length scale, $k_{\rm f}^{-1}$, to 
$C_{\rm S}\Delta$, which is the scale below which the eddy viscosity strongly damps motion. The effect of molecular viscosity is represented by the geometric
Reynolds number $Re$.
In the following we will take the limit of the latter to infinity or, equivalently, of vanishing molecular viscosity, and use $R$ as the control parameter. 
We will refer to the resulting system as governing an {\em LES fluid}. The other limit, in which $C_{\rm S}=0$ and $\nu>0$, will be referred
to as the DNS case.

Two quantities important for the analysis of the flow are the energy and enstrophy, which can be considered as the norm of velocity and vorticity, 
$\bm{\omega}=\nabla \times \bm{u}$, respectively:
\begin{alignat}{2}\label{en_ens}
E&=\frac{1}{L^3}\int \frac{1}{2}\|\bm{u}\|^2\,\mbox{d}\bm{x} &\qquad Q&=\frac{1}{L^3}\int \frac{1}{2}\|\bm{\omega}\|^2\,\mbox{d}\bm{x}
\end{alignat}
where the integral is taken over the entire domain and $\|.\|$ denotes the standard vector norm. In addition, we compute the mean rate of energy input,
\begin{equation}\label{EIR}
e=\frac{1}{\alpha L^3} \left<\int \mathbf{u}\!\cdot\!\mathbf{f}\,\mbox{d}\bm{x}\right>
\end{equation}
and transfer to sub-filter scales,
\begin{equation}\label{EDR}
\epsilon=\left<\frac{\mbox{d}E}{\mbox{d}t}\right>=2\left<\int \nu_{T}S_{ij}S_{ij}\,\mbox{d}\bm{x}\right>
\end{equation}
where $<.>$ denotes the time or ensemble average. 

\subsection{Symmetries}\label{symm}

It is straightforward to verify that equation \rf{nondimLES} is equivariant under a group of symmetries generated by the following transformations:
\begin{itemize}
\item Translation over any distance $d$ in the vertical direction, $T_{d}$.
\item Reflection in the $x$-direction, $S_x$.
\item Reflection in the $y$-direction, $S_y$.
\item Reflection in the $z$-direction, $S_z$.
\item Rotation about the axis $x=y=0$ over $\pi/2$ followed by a shift over $L/2$ in the x-direction, $R$.
\item A shift over $L/2$ in both the $x$ and $y$ directions, $D$.
\end{itemize}
In addition, we define the shift in time along a periodic orbit of period $P$ as
$Q_{\delta}$ where $0 \leq \delta<P$. We will not consider Galilean boosts and in the simulation code the flux in the vertical direction is identically equal to zero.

\subsection{Numerical methods}

For the purpose of numerical simulation the LES system is written on the Fourier basis
\begin{equation}\label{SDFT}
\bm{u}=\sum\limits_{\bm k} \hat{\bm{u}}(\bm{k}) e^{i \bm{k}\cdot\bm{x}}
\end{equation}
where each sum is taken from $-n/2+1$ to $n/2$ for a resolution of $n^3$ grid points. The pressure is eliminated by formulating the system in terms of
the vorticity. Two components of vorticity are time-stepped with a pseudo-spectral code using a
fourth-order accurate Runge-Kutta-Gill scheme with a time step of $0.05$ in non dimensional units. For de-aliasing we use
the phase-shift method introduced by Patterson and Orszag\cite{PatOrs}. The largest resolved wave number is $2\sqrt{2}/3\times n/2\approx 0.94 \times n/2$, where $n/2$ is the Nyquist wave number.
The total number of degrees of freedom
in a simulation is $28,484$ for $n=32$; $230,240$ for $n=64$ and $1,839,283$ for $n=128$.
The exact same methods are used to propagate perturbations to the vorticity field and the Smagorinsky parameter using the tangent linear model.

Using the time-steppers for the vorticity and its linear perturbations we can solve boundary value problems in time to find equilibria and periodic orbits
relative to the shift $T_{d}$. For this end, we use Newton-Krylov continuation. An review of this method was recently presented by S\'anchez and Net\cite{SanchNet},
and we follow their notes, using, in addition, Viswanath's method to handle the translation symmetry \cite{Vis}.

\section{The Smagorinksy parameter}\label{Smag_section}

The closure described above introduces a dimensionless parameter, $C_{\rm S}$, the value of which cannot be determined from modelling considerations alone.
In fact, it is not known what spatial filter $G$ gives rise to the LES fluid dynamics that system \rf{nondimLES} produces. In practice, the Smagorinsky parameter
is fixed by comparing statistical properties of the LES fluid to those of fully developed turbulence.
The central quantity used in the comparison 
is the energy spectrum. It describes how the energy is distributed over spatial scales and is defined in terms of the Fourier transform of velocity as
\begin{equation}\label{3Dspec}
E(k)\equiv \sum\limits_{k-\frac{1}{2}\leq \|\vec{k}\|<k+\frac{1}{2}} \frac{1}{2}<\|\hat{\bm{u}}(\vec{k})\|^2>
\end{equation}
We expect to see three distinct domains in the energy spectrum. In the {\em energy-containing subrange}, where $k\approx k_{\rm f}$, the flow is strongly influenced by 
the external forcing and the boundary conditions.
On the other end of the scale, when $k\approx 1/(c_{\rm S}\Delta)$ or larger, the eddy viscosity term is dominant. The domain in between these two extremes
is called the {\em inertial subrange}. Our interest in LES is mostly focused on the dynamics of fluid motion in the energy containing and inertial subrange.

\begin{figure}
\begin{center}
\resizebox{0.45\columnwidth}{!}{\includegraphics{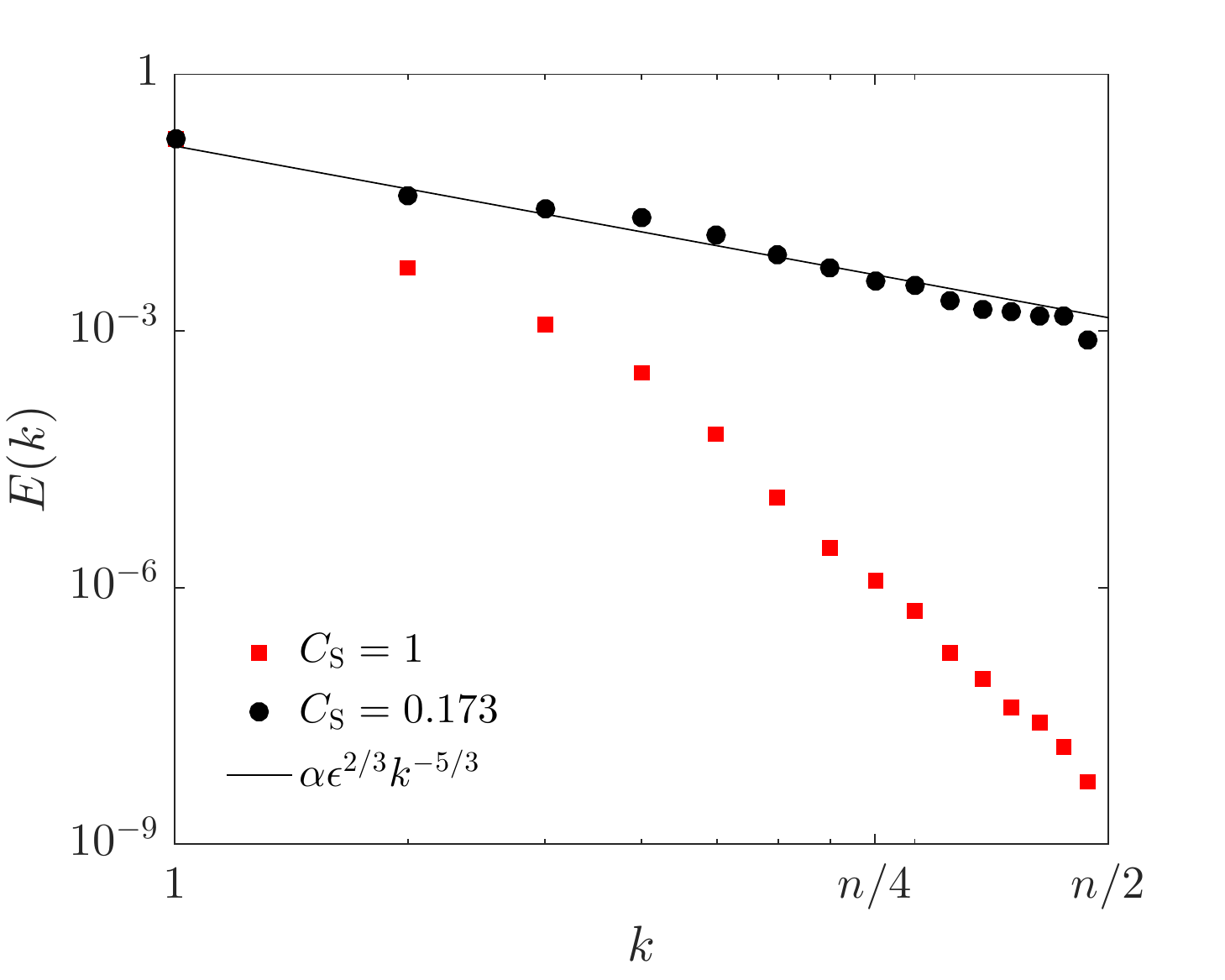} }\resizebox{0.45\columnwidth}{!}{\includegraphics{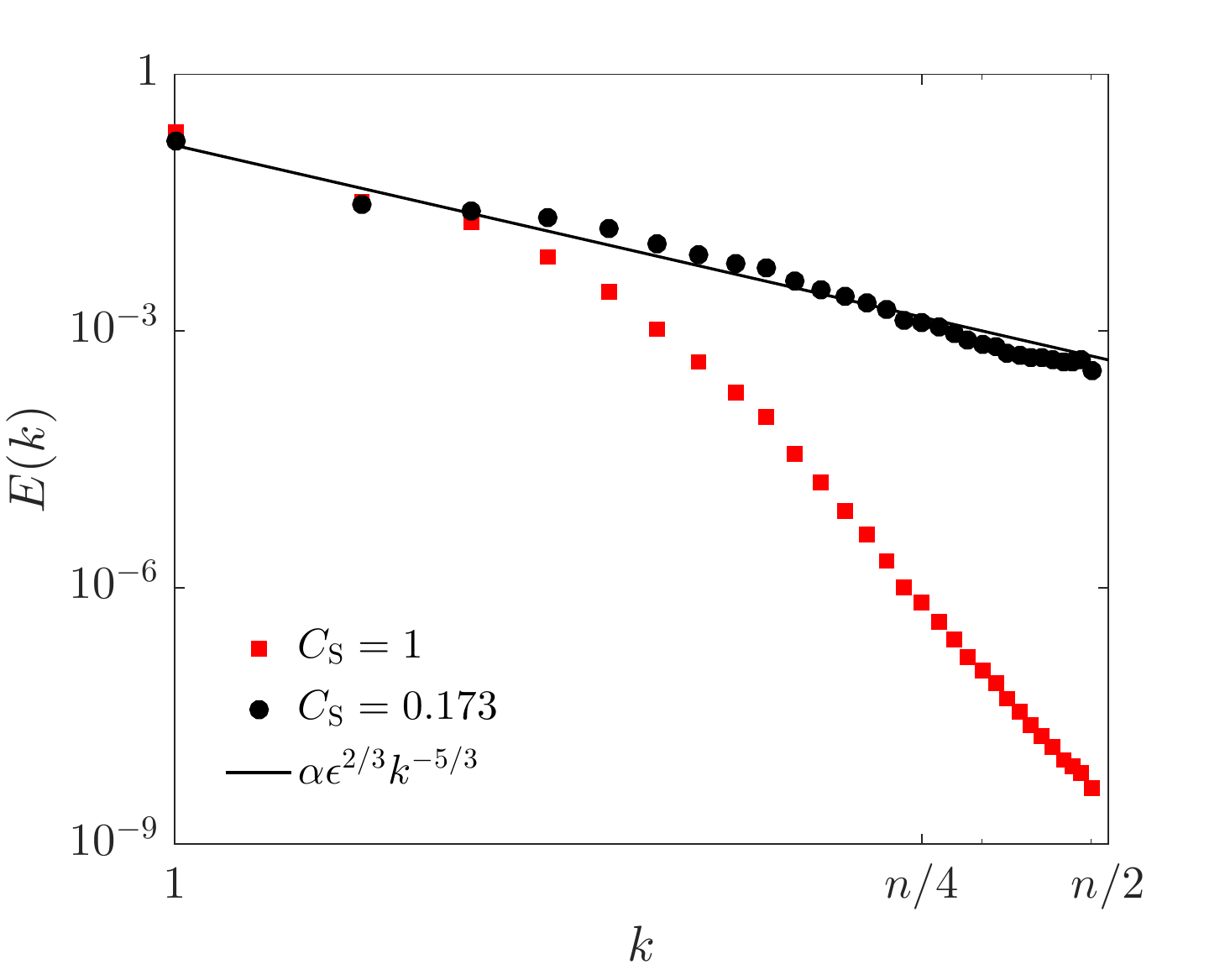} }
\end{center}
\caption{Energy spectrum of the LES fluid for $C_{\rm S}=1$ and $C_{\rm S}=0.173$ at resolutions $n=32$ (left) and $n=64$ (right). The straight lines denote the Kolmogorov spectrum
\rf{kol_spec} with $\alpha=1.5$ and $\epsilon$ as computed using expression \rf{EDR}. For $n=32$ we have $R=8.77$ and $R=91.0$ and for $n=64$ we have $R=22.1$ and $R=229$, respectively.}
\label{Spectra_1}
\end{figure}
For the DNS case, Kolmogorov formulated a theory for the energy spectrum in the inertial subrange \cite{Kolmo}. Under the assumption that the fluid 
motion on those scales is homogeneous and isotropic, and that its statistical properties are uniquely determined by the energy dissipation rate, the 
spectrum should take the form 
\begin{equation}\label{kol_spec}
E(k)=\alpha \epsilon^{2/3}k^{-5/3}
\end{equation}
where $\alpha$ is a parameter with some degree of universality, the value of which is usually taken to be $\alpha=1.5$. This spectrum is often called 
``the Kolmogorov spectrum'' or ``the $-5/3$ spectrum''. Various authors have attempted to pick a value for $C_{\rm S}$ that results in the spectrum of the
LES fluid resembling expression \rf{kol_spec} as closely as possible.

Lilly\cite{Lilly} derived a value for $C_{\rm S}$ under the assumptions that the kernel $G$ corresponds to a sharp cut-off filter in Fourier space, 
eliminating all Fourier coefficients beyond wave number $2\pi/l_{\rm f}=\pi/\Delta$. Fixing $\Delta=2\pi/n$, i.e. the grid spacing, he found that $C_{\rm S}=(2/(3\alpha))^{3/4}/\pi$
is consistent with a $-5/3$ spectrum for all wave numbers. Setting $\alpha=1.5$ gives $C_{\rm Lilly}\approx 0.173$,
the ``Lilly value''.  Muschinski\cite{Musch} used similarity hypotheses similar to those Kolmogorov made for Navier-Stokes flow to investigate the LES spectrum.
He concluded that the spatial filter that corresponds to the Smagorinsky model is most likely smoother than a sharp cut-off. Using more sophisticated forms
of the target energy spectrum at high wave numbers, he defined an effective filter length based on the wave number at which a significant deviation
from the Kolmogorov spectrum occurs. This filter length is several times larger than $C_{\rm S}\Delta$. Meyers and Sagaut \cite{MeySag} systematically studied 
target spectra and their corresponding optimal value for the Smagorinsky parameter. They found that the optimal value sensitively depends on the local
rate of energy dissipation on scales below $C_{\rm S}\Delta$ and that fixing the parameter to the Lilly value, independent of time and space, leads to 
an overestimation of overall dissipation. 
In the current study we will nonetheless stick to the simplest model, with $C_{\rm S}$ constant in time and space. Rather than fixing it to a particular value,
we use $C_{\rm S}$ or, equivalently, $R$, as a control parameter. Figure \ref{Spectra_1} demonstrates the effect of varying $R$ on the measured energy spectrum.
For small $R$, the effective filter length is too close to the forcing length scale and the spectrum decays much faster than the $-5/3$ power law. When $C_{\rm s}$
is set to the Lilly value, the Kolmogorov spectrum is recovered approximately all the way up to the cut-off wave number.


\section{Transitions for a grid of $32^3$ points}
\label{sec:transition}

\begin{figure}[t]
\begin{center}
\resizebox{0.75\columnwidth}{!}{\includegraphics{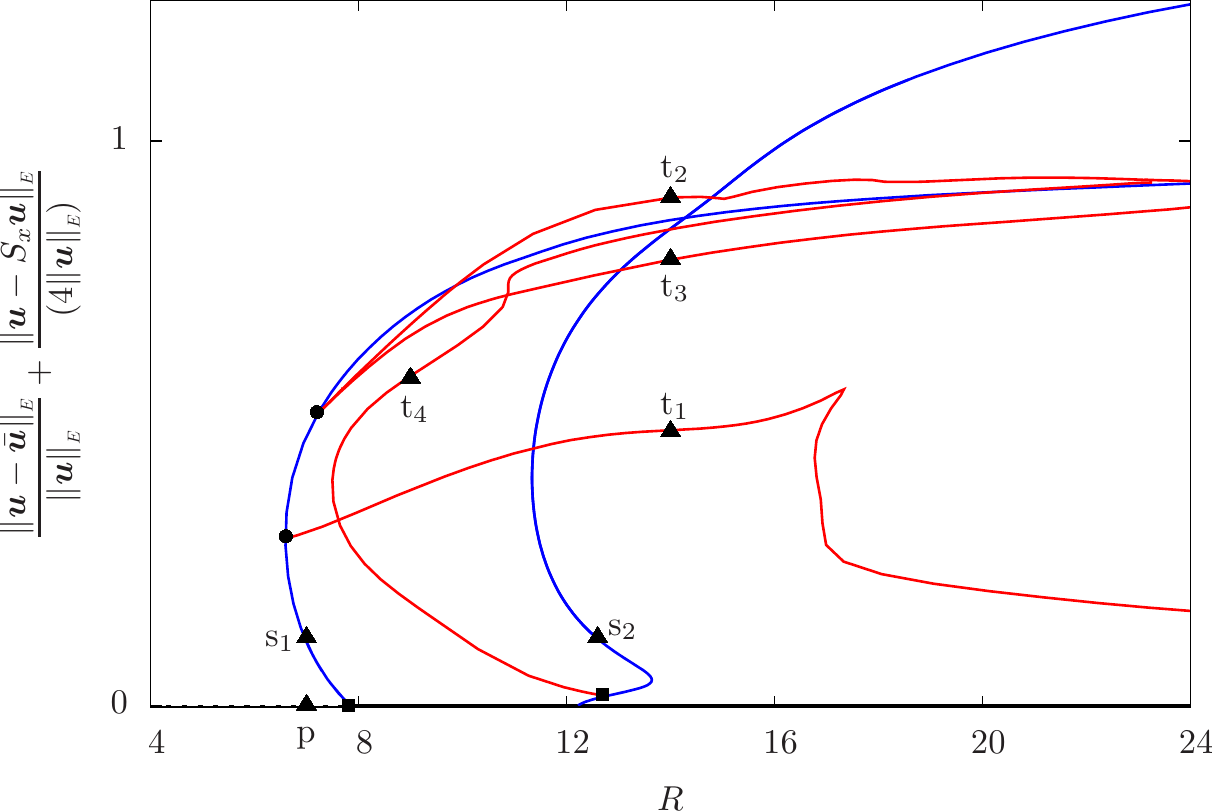} }
\end{center}
\caption{Partial bifurcation diagram of the LES fluid with Taylor-Green forcing. Shown on the vertical axis is the sum of the three-dimensional energy and the deviation from symmetry under the reflection $S_x$ as defined in the text. For periodic solutions, the sum of the maximal values over one period is shown. The black, dashed line denotes the stable primary equilibrium which loses stability at a Hopf bifurcation. All other branches are unstable. Secondary branches have been coloured blue and tertiary branches in red. Filled squares denote Hopf bifurcations, filled circles denote (equivariant) pitchfork bifurcations and filled triangles correspond to the physical space portraits \ref{prim}--\rf{ter4}.}
\label{partial}
\end{figure}

A partial bifurcation diagram of the LES fluid is shown in figure \ref{partial}. In order to differentiate between various types of solutions, we computed the energy norm, defined
in equation \rf{en_ens}, 
of each solution minus its vertical average,
\begin{equation}
\bar{\bm{u}}(x,y)=\frac{1}{L}\int\limits_{z=0}^{L}\bm{u}(x,y,z)\,\mbox{d}z 
\end{equation}
 and of its deviation from reflection symmetry in the $y$-$z$ plane, i.e. $\bm{u}-S_x\bm{u}$. The sum of the two measures, normalised by the maximum they can attain, is shown on the vertical axis.

\begin{figure}[t]
\begin{center}
\resizebox{0.75\columnwidth}{!}{\includegraphics{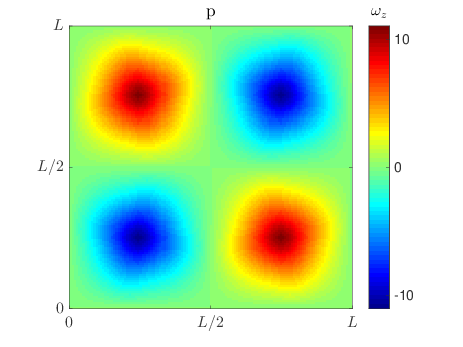} }
\end{center}
\caption{Physical space portrait of the viscous equilibrium. Shown is the vertical vorticity in a plane of constant height $z$. The label correspond to the one in figure \ref{partial}.}
\label{prim}
\end{figure}

For very small LES Reynolds numbers, an equilibrium that closely resembles the external forcing
appears to be the only invariant solution. We will refer to this family of solutions as the {\em primary branch} and to equilibria on this branch as the {\em viscous equilibria}.
In figure \ref{prim} the viscous equilibrium is shown at $R=7$. Since it is two-dimensional, i.e. invariant under translations along the $z$-axis, we plotted the vertical 
vorticity in a $x$-$y$ plane. The viscous equilibrium in invariant under all symmetries listed in section \ref{symm}.

\begin{figure}[t]
\begin{center}
\resizebox{0.49\columnwidth}{!}{\includegraphics{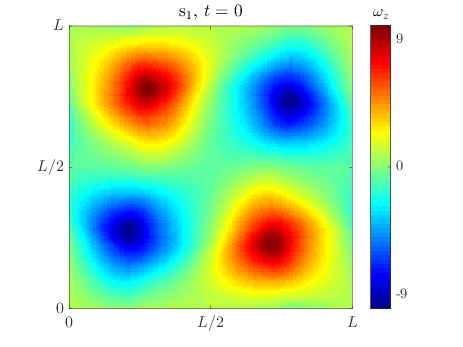} }\resizebox{0.49\columnwidth}{!}{\includegraphics{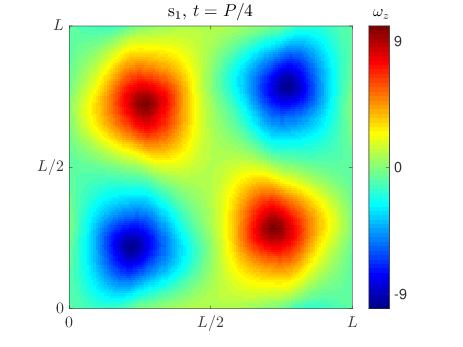} }
\end{center}
\caption{Physical space portraits of the secondary branch consisting of two-dimensional periodic orbits. Shown is the vertical vorticity in a plane of constant height $z$. Two snap shots are shown, separated by a quarter of the period. The label corresponds to the one in figure \ref{partial}.}
\label{sec1}
\end{figure}

As we increase the LES Reynolds number, the primary branch undergoes first a Hopf, around $R=7.84$, and then a pitchfork bifurcation around $R=12.2$. The former bifurcation gives rise to a branch
of two-dimensional periodic orbits. These orbits are also invariant under the symmetries mentioned in section \ref{symm} combined with a shift in time over part of the period. To be precise, the isotropy
subgroup of this branch of solutions is generated by $T_{d}$, $S_z$, $S_x\circ Q_{P/2}$, $S_y\circ Q_{P/2}$, $R \circ Q_{P/4}$ and $D \circ Q_{P/2}$. Two snap shots on a representative periodic orbit are
shown in figure \ref{sec1}. The vortex column have been deformed and rotate around the centres at $(\pi\pm \pi/2,\,\pi\pm \pi/2)$.

\begin{figure}[t]
\begin{center}
\resizebox{0.85\columnwidth}{!}{\includegraphics{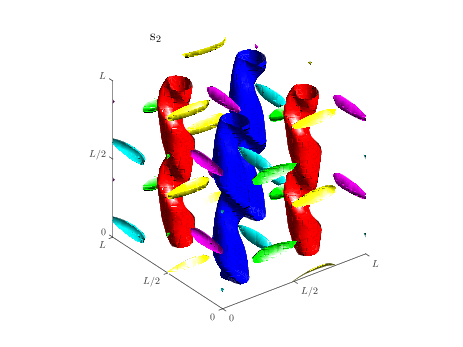} }
\end{center}
\caption{Physical space portrait of a solution on the second secondary branch consisting of three-dimensional equilibria. The label corresponds to the one in figure \ref{partial}. Shown are the isosurfaces of $z$-vorticity (red and blue for positive and negative), $x$-vorticity (green and yellow for positive and negative) and $y$-vorticity (cyan and magenta  for positive and negative) at $70\%$ of the maximal and minimal values.}
\label{sec2}
\end{figure}

The other secondary branch consists of three-dimensional equilibria. A representative solution is visualised in figure \ref{sec2} by means of isosurfaces of the vorticity. The large-scale vortices
are bent into a corkscrew-like shape with dominant wave number two in the vertical direction. In addition, pairs of smaller, counter-rotating vortices have appeared in the perpendicular directions.
The latter are reminiscent of so-called {\em ribs} that have previously been identified in Taylor-Green flow as a consequence of an instability concentrated at the hyperbolic stagnation points $(\pi/4 \pm \pi/4,\,\pi/4 \pm \pi/4)$ \cite{Leblanc}. The isotropy subgroup of the three-dimensional equilibria is
generated by $S_x\circ S_y \circ T_{L/4}$, $D$ and $R\circ T_{L/8}$.

\begin{figure}[t]
\begin{center}
\resizebox{0.85\columnwidth}{!}{\includegraphics{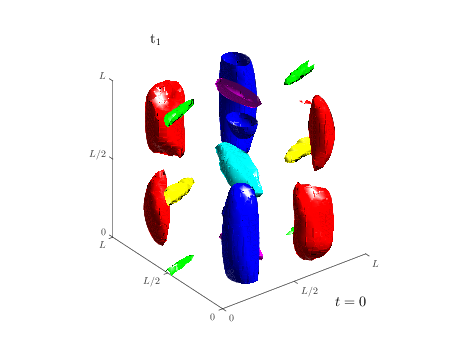} }\\
\resizebox{0.85\columnwidth}{!}{\includegraphics{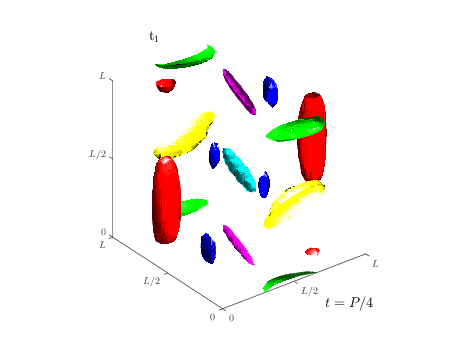} }
\end{center}
\caption{Physical space portrait of a solution on tertiary branch $\text{t}_1$, consisting of three-dimensional periodic orbits. The label corresponds to the one in figure \ref{partial}. Shown are two snap shots separated by one quarter of the period. The isosurfaces have been selected as coloured as in figure \ref{sec2}.}
\label{ter1}
\end{figure}

Both secondary branches give rise to several tertiary branches, of which we present four examples. The first originates from a pitchfork bifurcation of the secondary branch of
periodic orbits. Its symmetries are $S_z\circ S_y \circ S_x$, $D \circ Q_{P/2}$ and $T_{L/2}\circ  S_y \circ S_x$. Two snap shots of a representative solution are shown in figure \ref{ter1}.
Weak rib-like vortices are present in this solution, too, but perhaps its most striking feature is the weak projection on the external force. Particularly in the second snap shot
the four large-scale vortex columns are almost absent. Given that the separation between the largest and smallest scales at this LES Reynolds number is only a factor of seven, this
is surprising.

\begin{figure}[t]
\begin{center}
\resizebox{0.85\columnwidth}{!}{\includegraphics{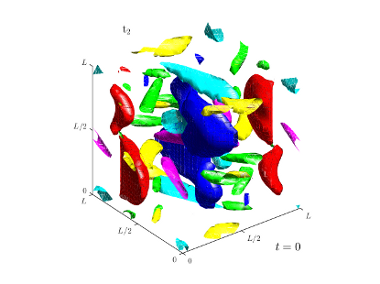} }\\
\resizebox{0.85\columnwidth}{!}{\includegraphics{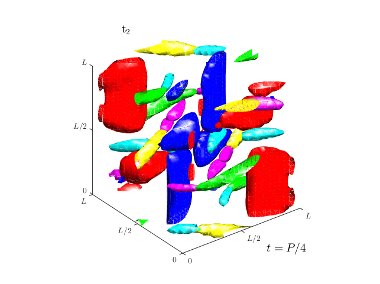} }
\end{center}
\caption{Physical space portrait of a solution on tertiary branch $\text{t}_2$, consisting of three-dimensional periodic orbits. The label corresponds to the one in figure \ref{partial}. Shown are two snap shots separated by one quarter of the period. The isosurfaces have been selected as coloured as in figure \ref{sec2}, except for the isolevels for the horizontal vorticity components, that have been fixed to $80\%$ of their maximal and minimal value.}
\label{ter2}
\end{figure}

\begin{figure}[t]
\begin{center}
\resizebox{0.85\columnwidth}{!}{\includegraphics{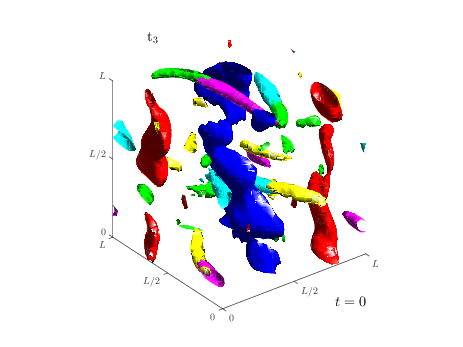} }\\
\resizebox{0.85\columnwidth}{!}{\includegraphics{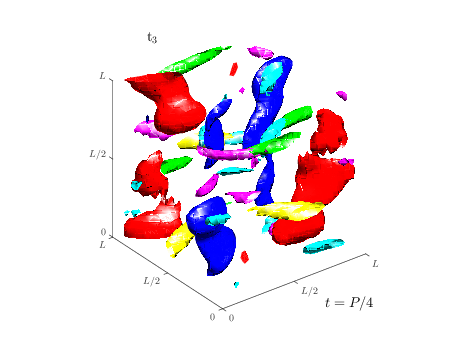} }
\end{center}
\caption{Physical space portrait of a solution on tertiary branch $\text{t}_3$, consisting of three-dimensional periodic orbits. The label corresponds to the one in figure \ref{partial}. Shown are two snap shots separated by one quarter of the period. The isosurfaces have been selected as coloured as in figure \ref{sec2}, except for the isolevels for the horizontal vorticity components, that have been fixed to $80\%$ of their maximal and minimal value.}
\label{ter3}
\end{figure}

Another two tertiary branches appear to originate from an equivariant pitchfork bifurcation. Numerical computation of the Floquet multipliers indicates that there are at least four 
nontrivial multipliers equal to unity. The symmetries
of the branches we computed are $S_z\circ S_y \circ S_x$, $D Q_{P/2}$, $R\circ R\circ Q_{P/2}$, $S_y \circ Q_{P/2}$ and $R\circ Q_{P/4}$, $D\circ Q_{P/2}$, $S_x\circ S_z\circ Q_{P/2}$. 
The first is portrayed in figure \ref{ter2}. In this solution, co-rotating pairs of large-scale vortices periodically approach the lines
$(\pi/4\pm\pi/4,\,\pi/4\pm\pi/4)$, creating strong strain fields in which smaller scale vortices are stretched. Similar dynamics can be observed in the second of the two, shown in figure
\ref{ter3}, but in this solution the horizontal components of vorticity are rather weak.

\begin{figure}[t]
\begin{center}
\resizebox{0.85\columnwidth}{!}{\includegraphics{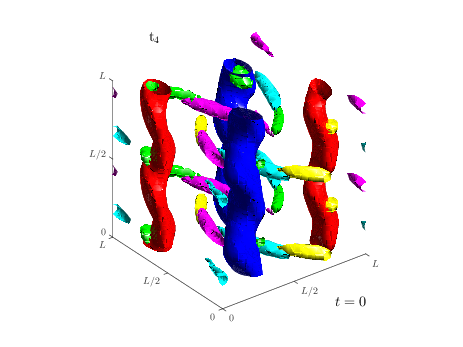} }\\
\resizebox{0.85\columnwidth}{!}{\includegraphics{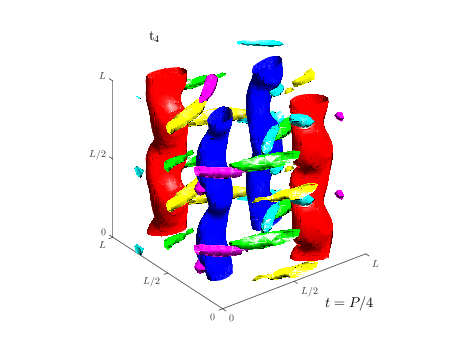} }
\end{center}
\caption{Physical space portrait of a solution on tertiary branch $\text{t}_4$, consisting of three-dimensional periodic orbits. The label corresponds to the one in figure \ref{partial}. Shown are two snap shots separated by one quarter of the period. The isosurfaces have been selected as coloured as in figure \ref{sec2}, except for the isolevels for the horizontal vorticity components, that have been fixed to $80\%$ of their maximal and minimal value.}
\label{ter4}
\end{figure}

Finally, we
describe the periodic orbits branching off the secondary branch of three-dimensional equilibria. Two snap shots of a representative solution are shown in figure \ref{ter4}. Its symmetries are
$S_x\circ S_y \circ T_{L/4}$, $D\circ Q_{P/2}$ and $R\circ T_{L/8} \circ Q_{P/4}$. In this solution, the rib-like vortices are generated alternatingly in the $x$ and $y$ directions.

\begin{figure}[t]
\begin{center}
\resizebox{0.65\columnwidth}{!}{\includegraphics{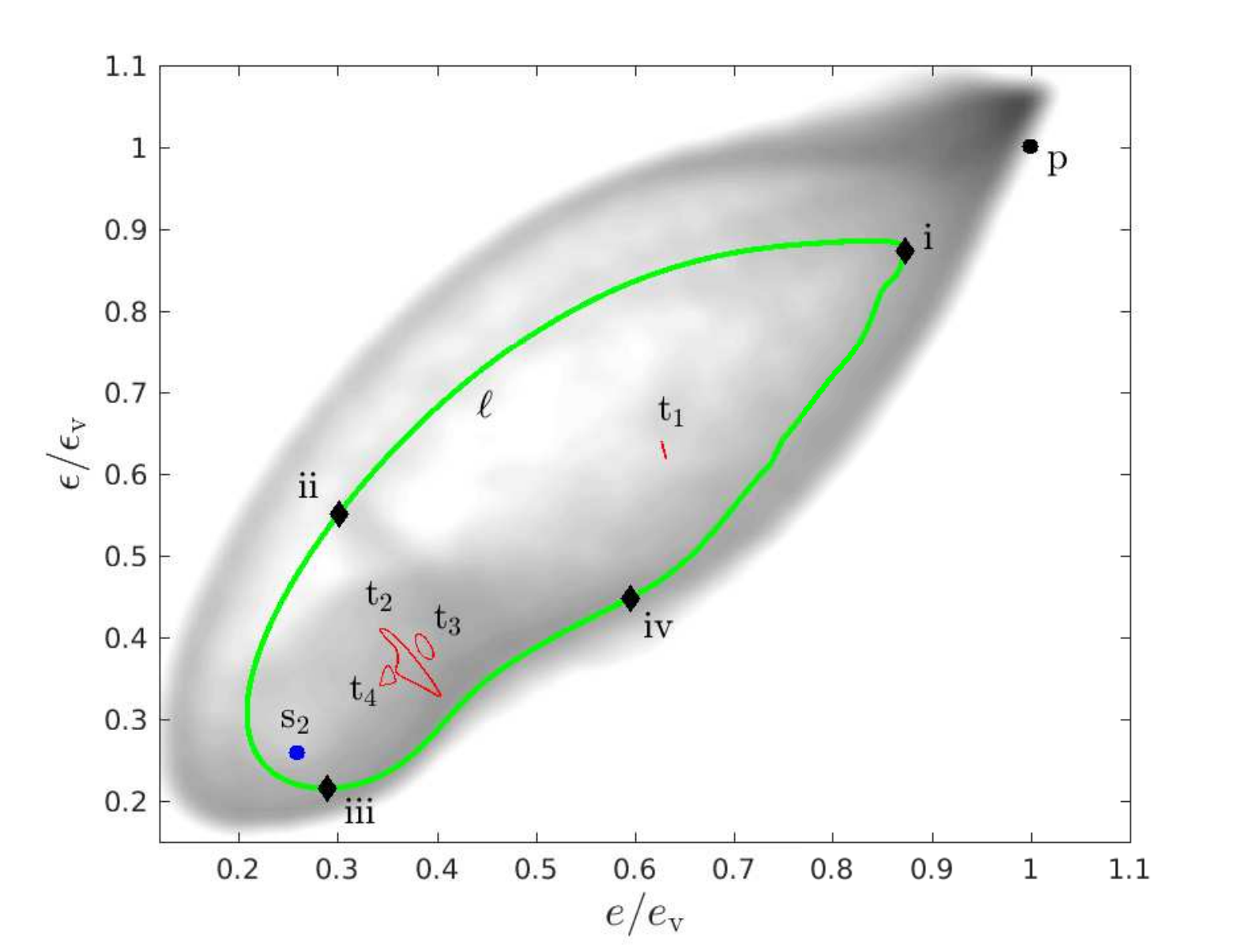} }
\end{center}
\caption{The PDF of weak turbulence and some of the invariant solutions presented in section \ref{sec:transition} projected on the rate of energy input and dissipation, normalised by the the rate found for the viscous equilibrium. The PDF was approximated by binning data points from a long, chaotic time series and is shown on a logarithmic grey scale. All data were obtained at $R=14$. The orbit shown in green, labeled $\ell$, was filtered from the chaotic time series by means of Newton-hook iteration. Physical space portraits along this orbit are shown in figure \ref{snapshots}.}
\label{EIR_EDR}
\end{figure}

Although there may exist tiny intervals in the LES Reynolds number for which some of the solutions described above are stable, the qualitative behaviour of the flow is mostly chaotic beyond the first Hopf bifurcation of the viscous equilibrium. Around $R=14$, the chaotic motion is rather structured. An example of such structured chaotic motion, or weak turbulence, is shown in figure \ref{EIR_EDR}. The Probability Density Function (PDF) of chaotic motion shows pathways for the flow to transition from the region near the viscous equilibrium to that near the secondary branch of equilibria and back. We have superimposed some of the the derivative solutions discussed above. Apart from the fact that they all have a number of discrete space-time symmetries, they also have a small amplitude compared to the chaotic motion. Although they exhibit some physically interesting properties, they do not resemble the unconstrained, weakly turbulent motion. 

\begin{figure}[t]
\begin{center}
\resizebox{0.49\columnwidth}{!}{\includegraphics{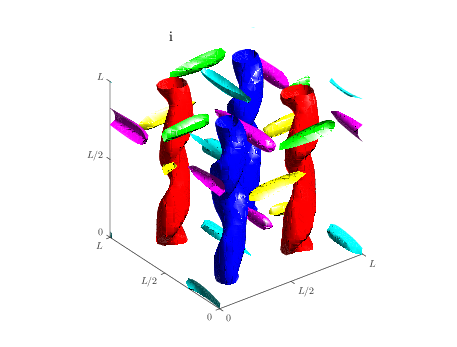} }\resizebox{0.49\columnwidth}{!}{\includegraphics{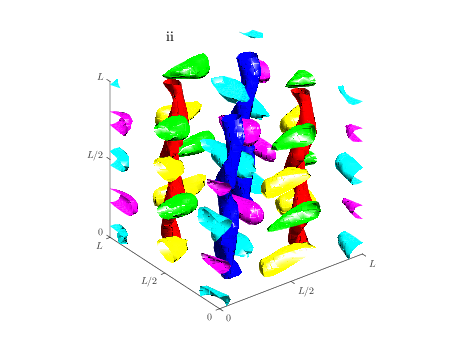} }
\resizebox{0.49\columnwidth}{!}{\includegraphics{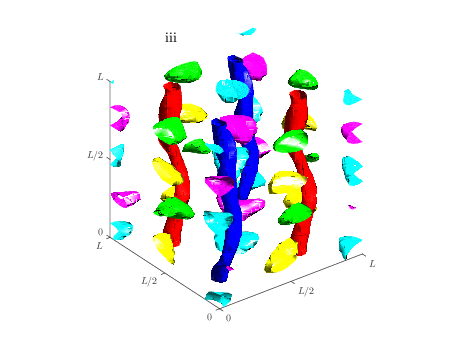} }\resizebox{0.49\columnwidth}{!}{\includegraphics{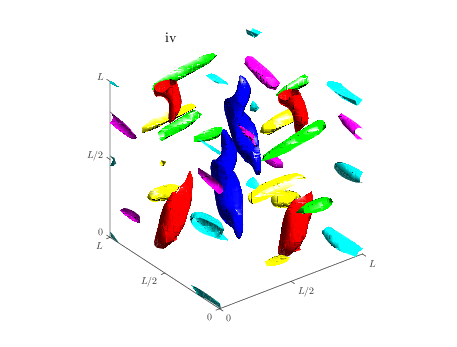} }
\end{center}
\caption{Four snapshots along the long-period, large-amplitude unstable periodic orbit labeled $\ell$ in figure \ref{EIR_EDR}. The time instants correspond to the black diamonds labeled with Roman numerals. At time $\mbox{i}$, the energy input and dissipation rate are close to their maximum along the orbit. }
\label{snapshots}
\end{figure}

We filtered one long-period, large-amplitude relative periodic orbit directly from the weak turbulence. In one period, the flow field shifts by approximately half the domain size in the vertical direction. Its projection on the rate of energy input and dissipation is shown in figure \ref{EIR_EDR} in green, while four snap shots are shown in figure \ref{snapshots}. This orbit does not have spatio-temporal symmetries other than vertical translations. 

Starting from label {\rm i} in figure \ref{EIR_EDR}, where the rate of energy input and dissipation are close to their maximum, we follow the periodic motion in the anti-clockwise
direction. At first, the flow field is a mild perturbation of the laminar flow, with pairs of weak horizontal vortices around the large-scale vortices induced by the external force.
The second snap shot is taken when the energy dissipation rate exceeds the input rate. At this time, the vertical vortices have weakened and more pairs of vortices have formed in the
perpendicular directions. In the third snap shot, close to the minimum of energy input and dissipation rate, the vertical vortices are still weak, and the secondary vortices have
also diminished in strength. In the final snap shot, where the rate of energy input exceeds that of dissipation, the vertical vortices have all but disappeared. From here, they
will slowly grow in strength under the influence of the external force. Although at this low LES Reynolds number we cannot expect to see detailed interaction across spatial scales, 
the evolution of the flow field hints at the existence of a regeneration cycle in which the large-scale vortices transfer energy to the smaller-scale ones before growing back in strength
under the influence of the forcing. We conjecture that solutions of this type may be able to reproduce turbulent statistics on larger grids and at smaller values of $C_{\rm s}$.

\section{Conclusion}

To the best of our knowledge, we have presented the first detailed analysis of the transition from laminar to weakly turbulent behaviour in LES on a three-dimensional, periodic domain.
We computed several derivative solutions, each with several discrete space-time symmetries and an amplitude small compared to that of the ambient turbulence. Although these solutions 
exhibit some physically interesting dynamics, they are not good models for the LES turbulence. A better candidate was obtained by filtering a relative periodic orbit directly
from a time series. This orbit has a large amplitude and along it the energy input rate, corresponding to the projection of the instantaneous flow field onto the external force,
shows great variation. We conjecture that on finer grids and for higher LES Reynolds numbers, this type of orbit is most likely to offer a good model for the unconstrained
turbulence.

The computation of the large-amplitude periodic orbit required several hundreds of Newton-hook iterations, each of which required around one hundred Krylov subspace iterations.
Clearly, it will be challenging to complete this computation in LES flows with a larger separation of length scales, for which the time-stepping and number of required Krylov
subspace iterations will be significantly higher. In ongoing work we are exploiting the acceleration offered by GPU computation to push the boundaries, ultimately aiming to 
detect inertial range dynamics in a SIS.

\section*{Acknowledgement}

This work was partially supported by a Grant-in-Aid for Scientific
Research (Grant Nos. 25249014, 26630055) from the Japanese Society for
Promotion of Science (JSPS). LvV was supported by an NSERC Discovery grant.


\begin{thebibliography}{}

\bibitem{KK}
G.~Kawahara and S.~Kida.
\newblock Periodic motion embedded in plane {C}ouette turbulence: regeneration
  cycle and burst.
\newblock {\em J. Fluid Mech.}, 449: 291-300, 2001.

\bibitem{VKK}
L.~van Veen, S.~Kida and G.~Kawahara.
\newblock Periodic motion representing isotropic turbulence.
\newblock {\em Fluid Dyn. Res.}, 38:19--46, 2006.

\bibitem{Halcrow}
J.~Halcrow, J.~F.~Gibson, P.~Cvitanovi\'c and V.~Viswanath.
\newblock Heteroclinic connections in plane Couette flow.
\newblock {\em J. Fluid Mech.}, 621: 365-376, 2009.

\bibitem{LK}
D.~Lucas and R.~R.~Kerswell.
\newblock Recurrent flow analysis in spatiotemporally chaotic 2-dimensional Kolmogorov flow.
\newblock {\em Phys. Fluids}, 27:045106, 2015.

\bibitem{POT}
P.~Cvitanovi\'c, R.~Artuso, R.~Mainieri, G.~Tanner and G.~Vattay.
\newblock {\em Chaos: Classical and Quantum},
{\tt ChaosBook.org} (Niels Bohr Institute, Copenhagen), 2016. 

\bibitem{SanchNet}
J.~S\'anchez Umbr\'ia and M.~Net
\newblock Numerical continuation methods for large--scale dissipative dynamical systems. 
\newblock {\em Eur. Phys. J.-Spec. Top.}, 225:2465, 2016.

\bibitem{OrsPat}
S.~A.~Orszag and G.~S.~Patterson.
\newblock Numerical simulation of three-dimensional homogeneous isotropic turbulence.
\newblock {\em Phys. Rev. Lett.}, 28:76, 1972,

\bibitem{CFMT}
P.~Constantin, P.~Foias, O.~P.~Manley and R.~Temam.
\newblock Determining modes and fractal dimension of turbulent flows.
\newblock {\em J. Fluid Mech.}, 150:427--440, 1985.

\bibitem{Sanch}
J.~S\'anchez Umbr\'ia, M.~Net, B.~Garcia-Archilla and C.~Sim\'o.
\newblock Newton–Krylov continuation of periodic orbits for Navier–Stokes flows.
\newblock {\em J. Comput. Phys.},  201:13--33, 2004.

\bibitem{Smagorinsky}
J.~Smagorinsky.
\newblock General circulation experiments witht the primitive equations. I. The basic experiment.
\newblock {\em Mon. Weather Rev.}, 91(3):99--164, 1963.

\bibitem{Lesieur}
M.~Lesieur and O.~M\'etais.
\newblock New trends in large eddy simulations of turbulence.
\newblock {\em Ann. Rev. Fluid Mech.}, 28:45--82, 1996.

\bibitem{TG}
G.~I. Taylor and A.~E. Green.
\newblock Mechanism of the production of small eddies from large ones.
\newblock {\em P. Roy. Soc. Lond. A Mat.}, 158(895):499--521, 1937.

\bibitem{Vis}
D.~Viswanath.
\newblock Recurrent motions within plane {C}ouette turbulence.
\newblock {\em J. Fluid Mech.}, 580:339--358, 2007.

\bibitem{PatOrs}
G.~S. Patterson Jr. and S.~.A. Orszag 
\newblock Spectral calculations of isotropic turbulence: efficient removal of aliasing interactions
\newblock {\em Phys. Fluids}, 14:2538--2541, 1971.

\bibitem{Kolmo}
A.~N.~Kolmogorov
\newblock The local structure of turbulence in incompressible viscous fluid for very large Reynolds numbers
\newblock {\em Dokl. Akad. Nauk SSSR}, 30:301--305, 1941 (English translation in {\em Proc. R.
Soc. Lond. A}, 434:9-13, 1991).

\bibitem{Lilly}
D.~K.~Lilly
\newblock The representation of small-scale turbulence in numerical simulation experiments
\newblock Technical report nr. 281, National Center for Atmospheric Research, Boulder Colorado, USA, 1966.

\bibitem{Musch}
A.~Muschinski.
\newblock A similarity theory of the locally homogeneous and isotropic turbulence generated by a Smagorinsky-type LES
\newblock {\em J. FLuid Mech.}, 225:239--260, 1996.

\bibitem{MeySag}
J.~Meyers and  P.~Sagaut
\newblock On the model coefficients for the standard and the variational multi-scale Smagorinsky model
\newblock {\em J. Fluid Mech.} 569:287--319, 2006.

\bibitem{Leblanc}
S.~Leblanc and F.~S.~Godeferd
\newblock  An illustration of the link between ribs and hyperbolic instability
\newblock {\em Phys. Fluids} 11(2):497--499, 1999.

\end{thebibliography}
\end{document}